# Shape of a sliding capillary contact

Valentin L. Popov

Technische Universität Berlin, Str. des 17. Juni 135, 10623 Berlin, Germany

*v.popov@tu-berlin.de*

**Abstract.** We consider a classical problem of a capillary neck between a parabolic body and a plane with a small amount of liquid in between. In the state of thermodynamic equilibrium, the contact area between the bodies and the liquid layer has a circular shape. However, if the bodies are forced to slowly move in the tangential direction, the shape will change due to the hysteresis of the contact angle. We discuss the form of the contact area under two limiting assumptions about the friction law in the boundary line.

**Keywords**: Capillarity, contact angle hysteresis, friction

## 1. Introduction

Since Coulomb, it is known that the force of dry friction may depend on humidity of the atmosphere [1],[2]. As a physical reason for this dependency, formation of capillary bridges is considered [3], because in a complete cycle of formation and destruction of a capillary bridge a certain amount of energy is dissipated. However, even if the capillary bridges are not destroyed, they can contribute to friction due to the dry friction experienced by the boundary line of a capillary bridge. The friction in the boundary line leads to the well-known effect of contact angle hysteresis [4]. As the contact angle determines the height of a capillary bridge, the contact hysteresis will lead to some deformation of the contact area. In the present paper we theoretically determine this deformation and estimate the force of friction due to the angle hysteresis.

## 2. Basic geometry of a capillary bridge

We observe a rigid sphere with radius $R$ in contact with a solid surface with a small amount of a liquid in between (Fig. 1).

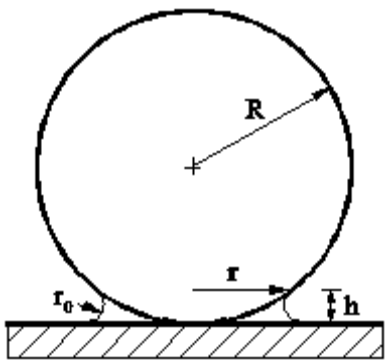

**Fig. 1** A capillary bridge between a rigid plane and a rigid sphere.

In equilibrium, the liquid forms a capillary bridge, which has two radii of curvature. The largest radius $r$ is always positive. The sign of the smaller radius depends on if the contact angle is larger or smaller than $\pi/2$. For small contact angles, in the case of wetting of the surface, $r_0$ is negative. There is a reduced pressure in the liquid, which leads to a force that we call capillary force. In the most cases one can assume $|r_0| \ll r$, the differences between the pressure inside and outside the contact can be written as [4]

$$\Delta p = -\frac{\gamma}{r_0}, \quad (1)$$

where $\gamma$ is the surface tension of the liquid. Note that in the equilibrium or at slow (quasistatic) sliding, the pressure in the liquid should be constant. This means, that the radius of curvature $r_0$ should be constant along the whole boundary line of liquid.

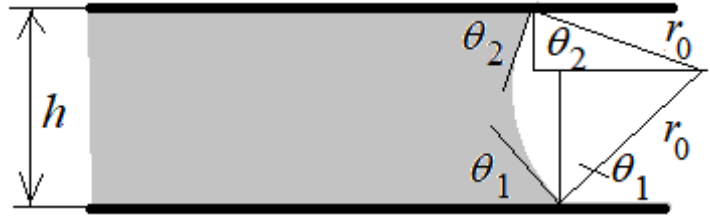

**Fig. 2 Geometry of a capillary bridge.**

Fig. 2 displays the geometry of the contact angles and meniscus of a capillary bridge. It can be easily seen that the following geometrical relation is valid:

$$r_0(\cos\theta_1 + \cos\theta_2) = h. \quad (2)$$

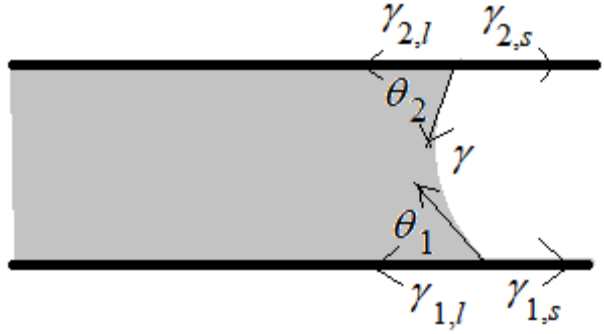

**Fig. 3 Scheme of tension forces acting at the boundary of a liquid bridge. $\gamma_{1,l}$ and $\gamma_{1,s}$ are the surface energies of the first solid in contact with liquid and the atmosphere, correspondingly. Similar notations are valid for the second body. $\gamma$ is the surface tension of liquid.**

The contact angles can be obtained from the equilibrium conditions for the boundary line, taking into account that each interface is acting on the boundary with a linear force density equal to the specific interface energy [4]:

$$\gamma_{1,s} = \gamma_{1,l} + \gamma\cos\theta_1, \quad (3)$$

$$\gamma_{2,s} = \gamma_{2,l} + \gamma\cos\theta_2, \quad (4)$$

which gives for the contact angles

$$\cos\theta_1 = \frac{\gamma_{1,s} - \gamma_{1,l}}{\gamma}, \quad (5)$$

$$\cos\theta_2 = \frac{\gamma_{2,s} - \gamma_{2,l}}{\gamma}. \quad (6)$$

Eq. (5) and (6) are valid in thermodynamic equilibrium. If the boundary of the bridge is moving, it can experience a force of dry friction similar to that described for adhesive contacts in [5]. In the present paper, we will for simplicity assume that such dry friction force does act only on the boundary of a fluid in contact with body 1. That means that the second body is considered as being very smooth and chemically homogeneous, so that the contact angle hysteresis on its surface can be neglected. In the presence of friction force, Eq. (3) will take the form

$$\gamma_{1,s} = \gamma_{1,l} + \gamma \cos\theta_1 \pm f, \qquad (7)$$

where $f$ is the linear density of the force of friction. The sing "+" should be used if the right boundary of liquid is moving to the right and "−" if it is moving to the left. For the left boundary, the use of sign is inverse to the above stated. From Eq. (7), it follows

$$\cos\theta_1 = \frac{\gamma_{1,s} - \gamma_{1,l} \mp f}{\gamma}. \qquad (8)$$

Substituting (1), (6) and (8) into (2) gives

$$h = \frac{\gamma_{1,s} + \gamma_{2,s} - \gamma_{1,l} - \gamma_{2,l} \mp f}{|\Delta p|} = \frac{\gamma^* \mp f}{|\Delta p|} \qquad (9)$$

with $\gamma^* = \gamma_{1,s} + \gamma_{2,s} - \gamma_{1,l} - \gamma_{2,l}$.

## 3. Shape of the contact area for friction perpendicular to the boundary line

The force of friction acting on the boundary line may depend on the nature of this force. Two limiting cases can be considered. If the microscopic heterogeneity is moderate, then the movement of the liquid in the direction of the boundary line will occur without instabilities. However, the appearance of instabilities is a necessary precondition for appearance of static force of friction [5]. This means, that the movement along the boundary does not lead to friction. The movement in the direction perpendicular to the boundary, on the contrary, will always lead to appearance of instable configurations followed by rapid jumps and energy dissipation. On the macroscopic scale, this energy dissipation is perceived as being due to static force of friction. Under the conditions described, the force of friction is always directed perpendicularly to the boundary, as shown in Fig. 4. Moreover, the magnitude of the force of friction does not depend on the velocity, so that the force line density is constant in all boundary points. However, its direction changes abruptly at the side points of the contact where the direction of movement is parallel to the boundary line.

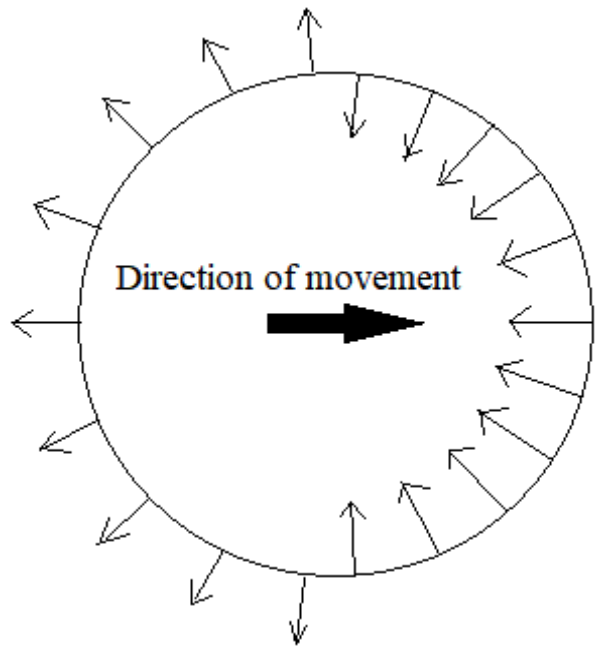

**Fig. 4 Friction forces acting on the boundary of liquid when the capillary contact is moving to the right.**

From Eq. (9), it follows that the height of the gap on the front side of contact is constant and equal to

$$h_{\text{front}} = \frac{\gamma^* - f}{|\Delta p|}, \qquad (10)$$

while at the back side it is

$$h_{\text{back}} = \frac{\gamma^* + f}{|\Delta p|}. \qquad (11)$$

For a parabolic body, this means that the contact area has constant (but different) radii on the front and back side of the contact, Fig. 5:

$$r_{0,\text{front}} = \sqrt{2R\frac{\gamma^* - f}{|\Delta p|}}, \quad r_{0,\text{back}} = \sqrt{2R\frac{\gamma^* + f}{|\Delta p|}}. \qquad (12)$$

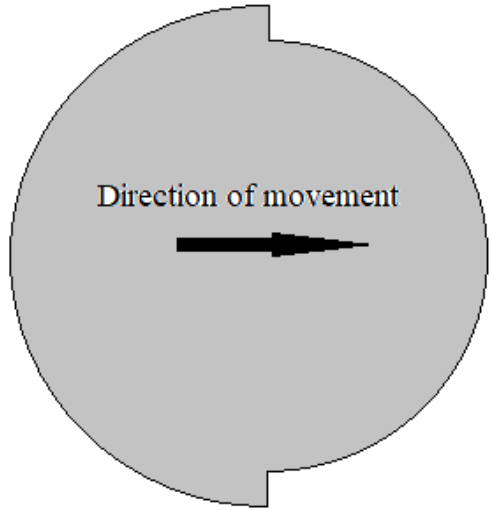

**Fig. 5 Shape of the contact area in the case of constant friction force directed perpendicular to the boundary of liquid.**

Note that the pressure difference $\Delta p$ can change during the movement, to maintain the given volume of interfacial liquid. However, the change of volume at a constant $\Delta p$ is zero in the linear approximation, so that if the force of friction is small, then the pressure difference can be considered as unchanged. In the same approximation, the total force of friction can be evaluated as

$$F_{\text{Friction}} = 4 \int_0^{\pi/2} Rf \cos\theta \, d\theta = 4Rf. \qquad (13)$$

## 4. Shape of the contact area for friction opposite to the direction of motion

Principally thinkable is also another law of friction for the boundary line. If the microscopic heterogeneity of the substrate becomes large, this will lead to comparable changes in the configuration of the boundary line, independently on the direction of its movement. This means, that the force of friction will be directed oppositely to the direction of motion, as shown in Fig. 6a. However, for the contact angle, only the component of the friction force perpendicular to the boundary line is relevant. This component is equal to

$$f_\perp = f \cos\varphi. \qquad (14)$$

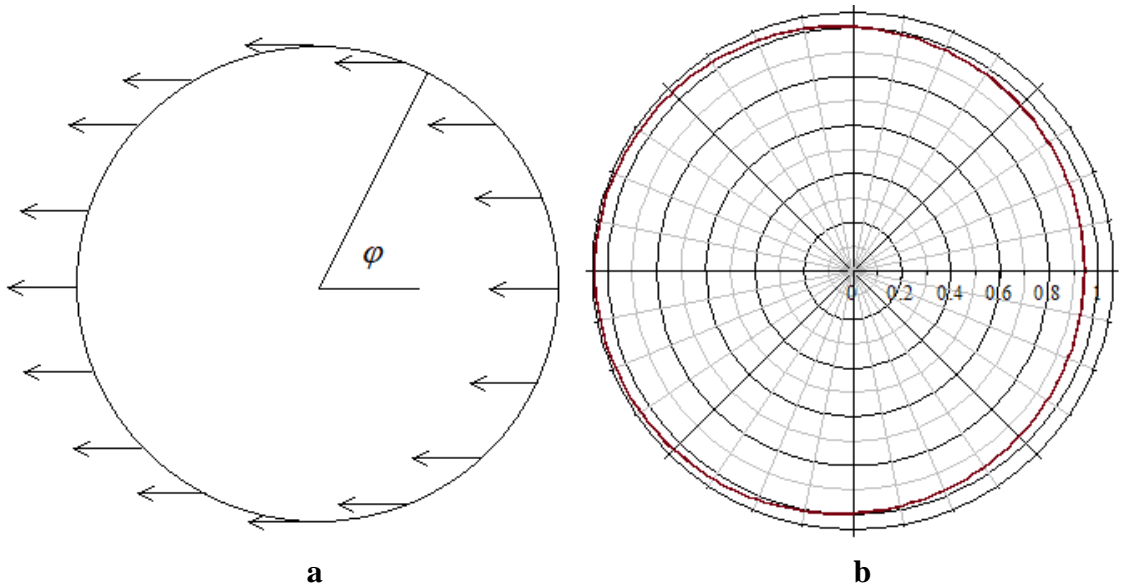

Fig. 6 Case of friction opposite to the direction of motion: (a) Forces acting on the boundary line, (b) shape of the contact area.

Eqs. (12) change to

$$r_0(\varphi) = \sqrt{2R\frac{\gamma^* - f\cos\varphi}{|\Delta p|}} \; . \tag{15}$$

This dependency is shown in Fig. 6b. While the shape of the contact is asymmetrical also in this case, the asymmetry is barely visible and manifests itself mostly in the shifting the contact area from the center of the body in the direction opposite to the direction of sliding.

## 5. Conclusions

We considered the changes in the shape of the contact area of a capillary bridge when it is sliding in tangential direction. Due to the friction force acting in the boundary, the contact area becomes asymmetric. Two limiting laws of friction have been considered: (a) with force of friction acting perpendicular to the boundary line and (b) opposite to the direction of sliding. The predicted shapes are substantially different. This means that experimental study of the shape of capillary bridges can give information about the character of the friction law in the boundary line.